\documentclass[twocolumn,showpacs,superscriptaddress,aps,prb]{revtex4-1}
\usepackage{bm}
\usepackage{amssymb,amsmath}
\usepackage[dvips]{graphicx}
\usepackage[colorlinks,linkcolor=blue,citecolor=blue,urlcolor=blue]{hyperref}

\begin{document}

\title{Percolation transition in quantum Ising and rotor models with sub-Ohmic dissipation}

\author{Manal Al-Ali}
\affiliation{Department of Physics, Missouri University of Science and Technology, Rolla, MO 65409, USA}

\author{Jos\'e A. Hoyos}
\affiliation{Instituto de F\'{\i}sica de S\~ao Carlos, Universidade de S\~ao Paulo,
C.P. 369, S\~ao Carlos, S\~ao Paulo 13560-970, Brazil}

\author{Thomas Vojta}
\affiliation{Department of Physics, Missouri University of Science and Technology, Rolla, MO 65409, USA}

\begin{abstract}
We investigate the influence of sub-Ohmic dissipation on randomly diluted quantum
Ising and rotor models. The dissipation causes the quantum dynamics of
sufficiently large percolation clusters to freeze completely. As a result, the zero-temperature
quantum phase transition across the lattice percolation threshold separates an unusual super-paramagnetic
cluster phase from an inhomogeneous ferromagnetic phase. We determine the low-temperature thermodynamic
behavior in both phases which is dominated by large frozen and slowly fluctuating percolation clusters.
We relate our results to the smeared transition scenario for disordered quantum phase transitions,
and we compare the cases of sub-Ohmic, Ohmic, and super-Ohmic dissipation.

\end{abstract}

\date{\today}

\maketitle

\section{Introduction}

The interplay between geometric, quantum, and thermal fluctuations in randomly diluted quantum
many-particle systems leads to a host of unconventional low-temperature phenomena. These include
the singular thermodynamic and transport properties in quantum Griffiths phases\cite{ThillHuse95,YoungRieger96}
as well as the exotic scaling behavior of the quantum phase transitions between different ground state
phases.\cite{Fisher92,Fisher95} Recent reviews of this topic can be found, e.g., in
Refs.\ \onlinecite{Vojta06,Vojta10}.

An especially interesting situation arises if a quantum many-particle system is diluted beyond the percolation
threshold $p_c$ of the underlying lattice (see, e.g., Ref.\ \onlinecite{VojtaHoyos08b} and references therein).
Although the resulting percolation quantum phase transition is driven by
the geometric fluctuations of the lattice, the quantum fluctuations lead to critical behavior different from that
of classical percolation. In the case of a diluted transverse-field Ising magnet, the transition displays exotic
activated  (exponential) dynamic scaling\cite{SenthilSachdev96} similar to what is observed at infinite-randomness
critical points.\cite{Fisher92,Fisher95} The percolation transition of the quantum rotor model shows conventional
scaling (at least in the particle-hole symmetric case where topological Berry phase terms are
unimportant\cite{FernandesSchmalian11}), but with critical exponents that differ from their classical
counterparts.\cite{VojtaSchmalian05,VojtaSknepnek06} For site-diluted Heisenberg quantum antiferromagnets,
 further modifications of the critical behavior were attributed to uncompensated geometric Berry phases.\cite{WangSandvik06,WangSandvik10}

In many realistic systems, the relevant degrees of freedom are coupled to an environment of ``heat-bath''
modes. The resulting dissipation can qualitatively change the low-energy properties of a quantum many-particle system.
In particular, it has been shown that dissipation can further enhance the effects of randomness on quantum
phase transitions. In generic random quantum Ising models, for instance, the presence of Ohmic dissipation completely
destroys the sharp quantum phase transition by smearing\cite{MillisMorrSchmalian01,Vojta03a,SchehrRieger06,SchehrRieger08,HoyosVojta08,HoyosVojta12}
while it leads to
infinite-randomness critical behavior in systems with continuous-symmetry order parameter.\cite{HoyosKotabageVojta07,DRMS08,VojtaKotabageHoyos09}
Interestingly, super-Ohmic dissipation does not change the universality class of random quantum Ising models~\cite{SchehrRieger08,HoyosVojta12}
 but plays a major role in systems with continuous-symmetry order parameter.~\cite{VojtaHoyosMohanNarayanan11}

It is therefore interesting to ask what are the effects of dissipation on randomly diluted quantum
many-particle systems close to the percolation threshold. It has recently been shown that Ohmic dissipation
in a diluted quantum Ising model leads to an unusual percolation quantum phase transition\cite{HoyosVojta06} at which
some observables show classical critical behavior while others are modified by quantum fluctuations.

In the present paper, we focus on the influence of sub-Ohmic dissipation (which is qualitatively stronger than the
more common Ohmic dissipation) on diluted quantum Ising models and quantum rotor models.
When coupled to a sub-Ohmic bath, even a single quantum spin displays a nontrivial
quantum phase transition from a fluctuating to a localized phase\cite{BullaTongVojta03} whose properties have attracted considerable attention recently
(see, e.g., Ref.\ \onlinecite{WRVB09} and references therein).
Accordingly, we find that the quantum dynamics of sufficiently large percolation clusters freezes completely as a result of
the coupling to the sub-Ohmic bath, effectively turning them into classical moments. The interplay between large
frozen clusters and smaller dynamic clusters gives rise to unconventional properties of the percolation transition
which we explore in detail.

Our paper is organized as follows: In Sec.\ \ref{sec:models}, we define our models and discuss their phase diagrams
at a qualitative level. Section \ref{sec:large-N} is devoted to a detailed analysis of the quantum rotor model in the
large-$N$ limit where all calculations can be performed explicitly. In Sec.\ \ref{sec:general}, we go beyond the
large-$N$  limit and develop a general scaling approach. We conclude in Sec.\ \ref{sec:conclusions}.

\section{Models and phase diagrams}
\label{sec:models}
\subsection{Diluted dissipative quantum Ising and rotor models}
\label{subsec:models}

We consider two models. The first model is a $d$-dimensional ($d\ge 2$) site-diluted transverse-field Ising
model\cite{Harris74b,Stinchcombe81,Santos82,SenthilSachdev96} given by the Hamiltonian
\begin{equation}
H_I= -J\sum_{\langle i,j\rangle
}\eta_{i}\eta_{j}\sigma_{i}^{z}\sigma_{j}^{z}-h_x \sum_{i}\eta_{i}\sigma_{i}^{x}~,
\label{eq:H}
\end{equation}
a prototypical disordered quantum magnet. The Pauli matrices $\sigma_{i}^{z}$ and
$\sigma_{i}^{x}$ represent the spin components at site $i$, the exchange interaction $J$
couples nearest neighbor sites, and the transverse field $h_x$ controls the quantum
fluctuations. Dilution is introduced via the random variables $\eta_{i}$ which can take
the values 0 and 1 with probabilities $p$ and $1-p$, respectively. We now couple each
spin to a local heat bath of harmonic
oscillators,\cite{CugliandoloLozanoLozza05,SchehrRieger06}
\begin{equation}
H = H_I + \sum_{i,n} \eta_i \left[\nu_{i,n}a_{i,n}^{\dagger}a_{i,n}^{\phantom{\dagger}}+ \frac 1 2
\lambda_{i,n}\sigma_{i}^{z} (a_{i,n}^{\dagger}+a_{i,n}^{\phantom{\dagger}}) \right],
\label{eq:Hamiltonian}
\end{equation}
where $a_{i,n}$  ($a_{i,n}^{\dagger}$) is the annihilation (creation) operator of
the $n$-th oscillator coupled to spin $i$; $\nu_{i,n}$ is its natural frequency, and
$\lambda_{i,n}$ is the coupling constant. All baths have the same spectral function
\begin{equation}
{\cal E}(\omega)=\pi \sum_{n}\lambda_{i,n}^{2} \delta (\omega-\nu_{i,n})=2\pi
\alpha\omega_c^{1-\zeta}\omega^\zeta e^{-\omega/\omega_{c}},
\label{eq:spectral_function}
\end{equation}
with $\alpha$ and $\omega_{c}$ being the dimensionless dissipation strength
and the cutoff energy, respectively. The exponent $\zeta$ characterizes the type of dissipation;
we are mostly interested in the sub-Ohmic case $0<\zeta<1$. For comparison, we will also consider
the Ohmic ($\zeta=1$) and super-Ohmic cases ($\zeta>1$). Experimentally, local dissipation (with various
spectral densities) can be realized, e.g., in molecular magnets weakly coupled
to nuclear spins\cite{ProkofevStamp00,CWMBB00} or in magnetic nanoparticles in an insulating host.
\cite{Wernsdorfer01}

The second model is a site-diluted dissipative quantum rotor model which can be conveniently defined
in terms of the effective Euclidean (imaginary time) action \cite{VojtaSchmalian05}
\begin{eqnarray}
\mathcal{A} &=&\int {\rm d}\tau \sum_{\langle ij\rangle }J\eta _{i}\eta
_{j}\bm{\phi}_{i}(\tau )\cdot \bm{\phi}_{j}(\tau )+\sum_{i}\eta _{i}%
\mathcal{A}_{\mathrm{dyn}}[\bm{\phi}_{i}]  \notag \\
\mathcal{A}_{\mathrm{dyn}}[\bm{\phi}] &=&\frac \alpha 2 \, T\sum_{\omega_n} \omega_c^{1-\zeta} |\omega
_{n}|^{\zeta}~ \tilde{\bm{\phi}} (\omega _{n})\cdot \tilde{\bm{\phi}} (-\omega _{n})~. \label{eq:rotoraction}
\end{eqnarray}%
Here, the random variables $\eta _{i}=0,1$ again implement the site
dilution, and $\omega_n$ are bosonic Matsubara frequencies. The rotor at site $i$ and imaginary time $\tau$ is described by
 $\bm{\phi}_{i}(\tau )$: a $N$-component vector of length $N^{1/2}$. Its Fourier transform in imaginary time is denoted by $\tilde{\bm{\phi}}(\omega _{n})$.
 The dynamic action
$\mathcal{A}_{\mathrm{dyn}}$ stems from integrating out the heat-bath modes, with the
parameter $\alpha$ measuring the strength of the dissipation, and the exponent $\zeta$
characterizing the type of the dissipation, as in the first model [see Eq.~(\ref{eq:spectral_function})].

\subsection{Classical percolation theory}

\begin{figure*}
\includegraphics[height=5cm]{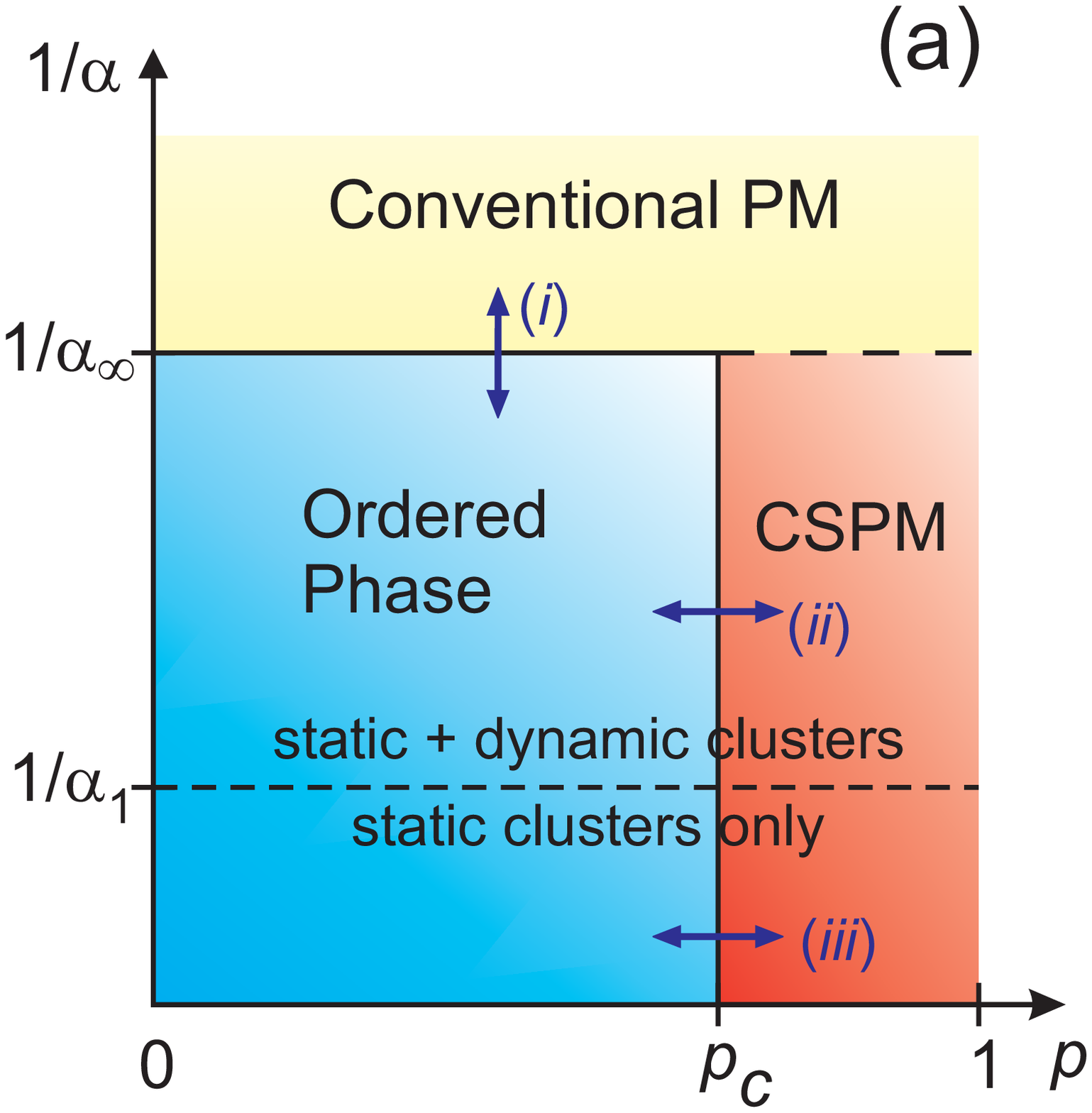}~\includegraphics[height=5cm]{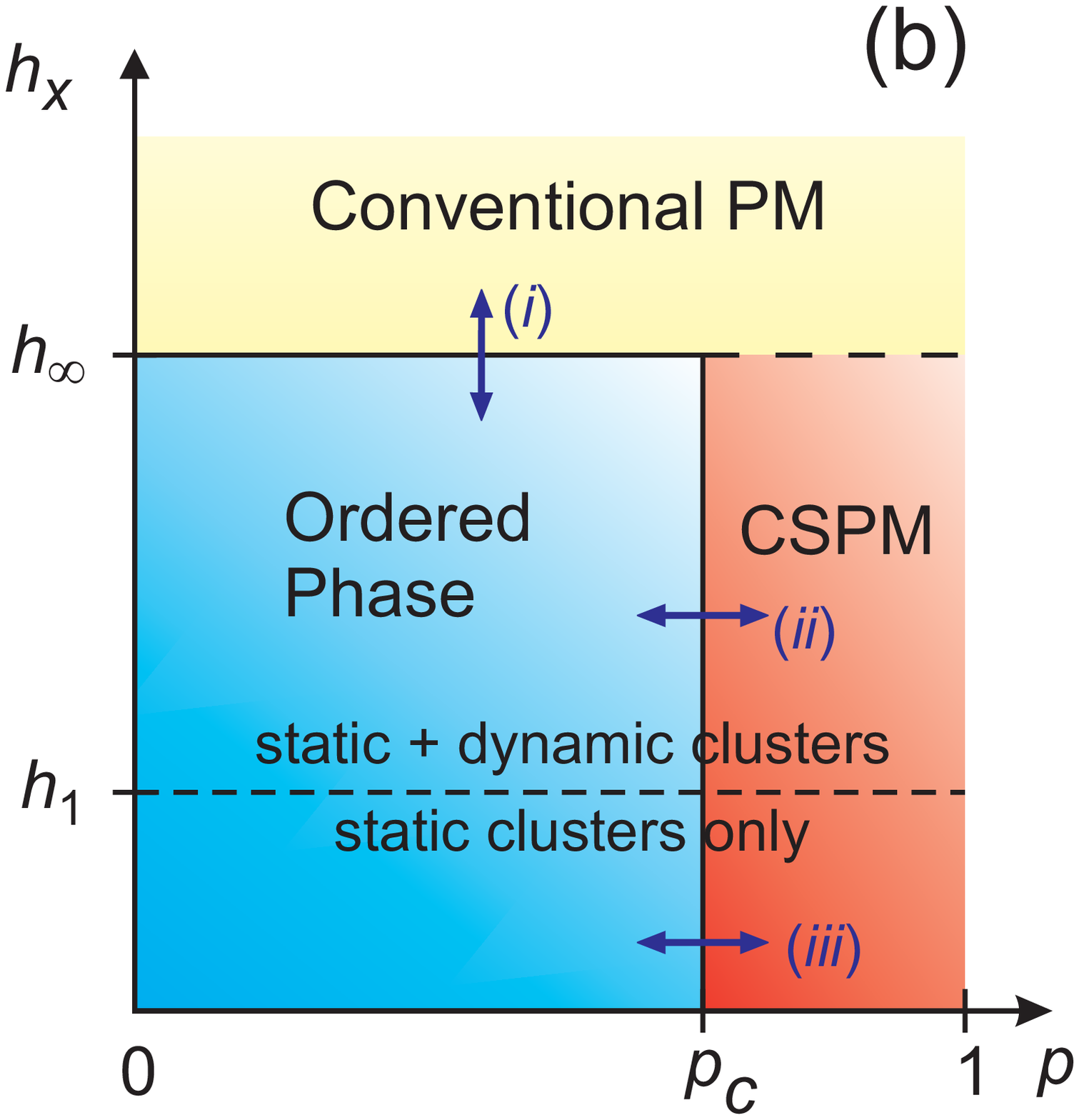}~\includegraphics[height=5cm]{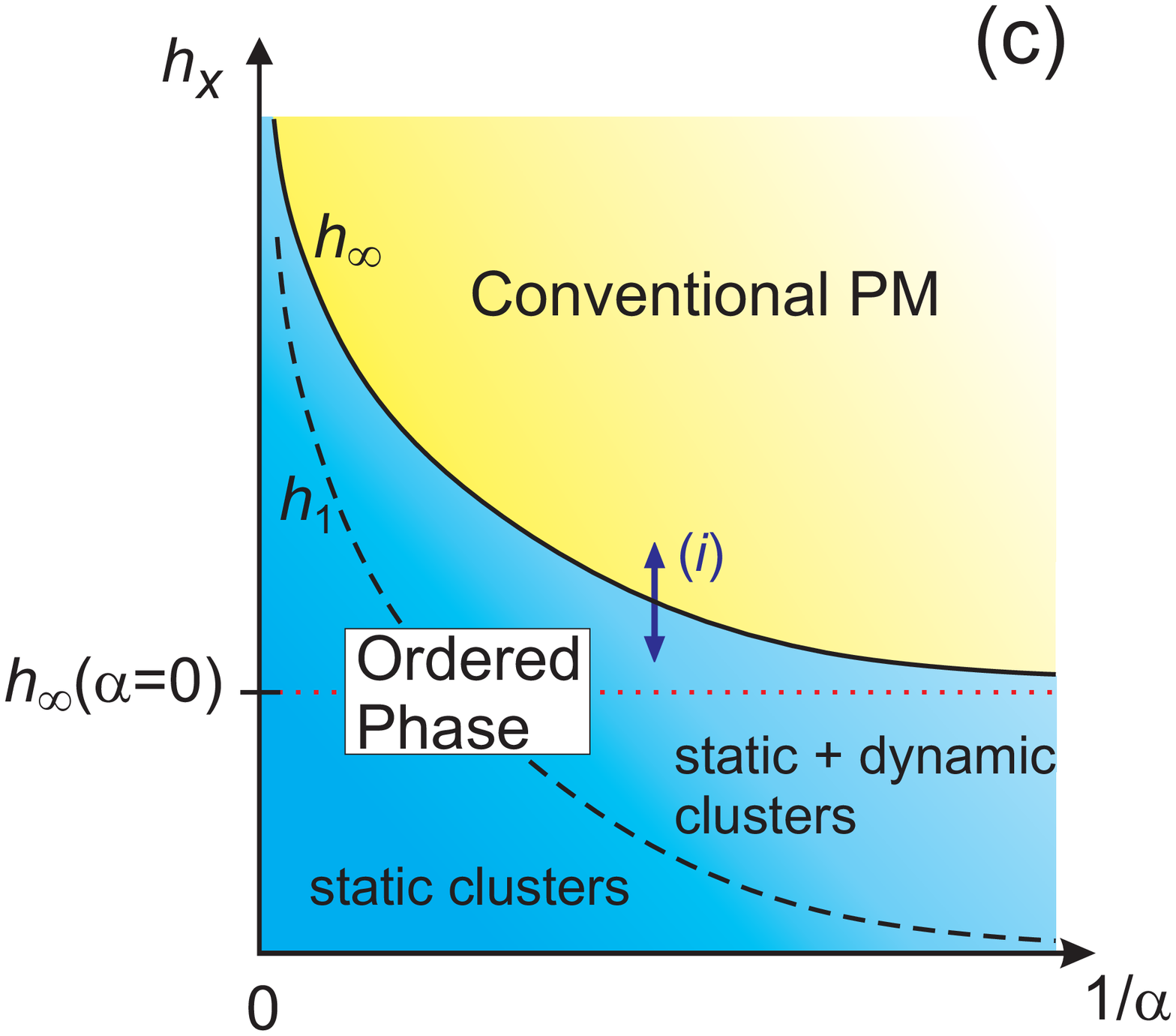}
\caption{(Color online) Schematic ground state phase diagram of the diluted dissipative quantum Ising model Eq.~(\ref{eq:Hamiltonian})
for fixed values of $\zeta<1$, $\omega_c$, and $J$. The three panels show three cuts through the three-dimensional
parameter space of dilution $p$, transverse field $h_x$, and dissipation strength $\alpha$. (a) $\alpha$--$p$ phase diagram
at a fixed transverse field $h_x$  with $h_x > h_\infty(\alpha=0)$ such that the dissipationless system is in the paramagnetic phase.
This phase diagram also applies to the rotor model Eq.~(\ref{eq:rotoraction}). (b) $h_x$--$p$ phase diagram
at a fixed dissipation strength $\alpha$. (c) $h_x$--$\alpha$ phase diagram at fixed dilution $p<p_c$. CSPM refers to the
cluster super-paramagnetic phase, transition (i) denotes the smeared generic (field or dissipation-driven) quantum phase transition,
and (ii) and (iii) denote the percolation quantum phase transitions in the two regimes with or without dynamic clusters, respectively.}
\label{fig:Ising_pd}
\end{figure*}

We now briefly summarize the results of percolation
theory\cite{StaufferAharony_book91} to the extent necessary for our purposes.

Consider a regular $d$-dimensional lattice in which each site is removed at random with
probability $p$.\footnote{In agreement with Subsec.\ \ref{subsec:models}, we define $p$ as the fraction of sites removed rather than
the fraction of sites present.} For small $p$, the resulting diluted lattice is
still connected in the sense that there is a cluster of connected
nearest neighbor sites (called the percolating cluster) that spans the entire system.
For large $p$, on the other hand, a percolating cluster does not exist.
Instead, the lattice is made up of many isolated clusters consisting of just a few sites.

In the thermodynamic limit of infinite system volume, the two regimes are separated by a sharp
geometric phase transition at the percolation threshold $p=p_c$.
The behavior of the lattice close to $p_c$
can be understood as a geometric critical phenomenon.
The order parameter is the probability $P_\infty$ of a site to belong to the infinite
connected percolation cluster. It is obviously zero in the disconnected phase ($p>p_c$)
and nonzero in the percolating phase ($p<p_c$). Close to $p_c$, it varies as
\begin{equation}
P_\infty \sim |p-p_c|^{\beta_c} \qquad (p<p_c)
\label{eq:percolation-beta}
\end{equation}
where $\beta_c$ is the order parameter critical exponent of classical percolation.
(We use a subscript $c$ to distinguish quantities associated with the
lattice percolation transition from those of the quantum phase transitions discussed
below). In addition to the infinite cluster, we also need to characterize the finite
clusters on both sides of the percolation threshold. Their typical size, the correlation or
connectedness length $\xi_c$, diverges as
\begin{equation}
\xi_c \sim |p-p_c|^{-\nu_c}
\label{eq:percolation-nu}
\end{equation}
with $\nu_c$ the correlation length exponent. The average mass $S_c$ (number of sites) of a
finite cluster diverges with the susceptibility exponent $\gamma_c$ according to
\begin{equation}
S_c \sim |p-p_c|^{-\gamma_c}~.
\label{eq:percolation-gamma}
\end{equation}

The complete information about the percolation critical behavior is contained in the cluster
size distribution $n_s$, i.e., the number of clusters with $s$ sites excluding the infinite
cluster (normalized by the total number of lattice sites). Close to the percolation
threshold, it obeys the scaling form
\begin{equation}
n_{s} (p) =s^{-\tau_c }f\left[ (p-p_c) s^{\sigma_c }\right] .
\label{eq:percscaling}
\end{equation}
Here, $\tau_c $ and $\sigma_c$ are critical exponents.
The scaling function $f(x)$ is analytic for small $x$ and has a single maximum at
some $x_{\rm max}>0$. For large $|x|$, it drops off rapidly
\begin{eqnarray}
f(x) &\sim&  \exp\left(- B_1 x^{1/\sigma_c}\right) \quad~~~~~~~~~~ ~(x>0),
\label{eq:scaling-function-disconnected}\\
f(x) &\sim&     \exp\left[- \left(B_2 x^{1/\sigma_c}\right)^{1-1/d}\right] \quad ~(x<0),
\label{eq:scaling-function-connected}
\end{eqnarray}
where $B_1$ and $B_2$ are constants of order unity. The classical percolation exponents are
determined by $\tau_c$ and $\sigma_c$: the correlation lengths exponent
$\nu_c =({\tau_c -1})/{(d\sigma_c )}$, the order parameter exponent
$\beta_c=(\tau_c-2)/\sigma_c$, and the susceptibility exponent
$\gamma_c=(3-\tau_c)/\sigma_c$.

Right at the percolation threshold, the cluster size distribution does not contain a
characteristic scale, $n_s\sim s^{-\tau_c}$, yielding a fractal critical percolation cluster
of fractal dimension $D_f=d/(\tau_c-1)$.

\subsection{Phase diagrams}
\label{subsec:PD}

Let us now discuss in a qualitative fashion the phase diagrams of the models introduced in Subsec.~\ref{subsec:models},
beginning with the diluted dissipative quantum Ising model Eq.~(\ref{eq:Hamiltonian}). If we fix the bath
parameters $\zeta$ and $\omega_c$ and measure all energies in terms of the exchange interaction $J$, we  still need
to explore the phases in the  three-dimensional parameter space of transverse field $h_x$, dissipation strength $\alpha$ and dilution $p$.
A sketch of the phase diagram is shown in Fig.\  \ref{fig:Ising_pd}.
For sufficiently large transverse field and/or sufficiently weak dissipation, the ground state is paramagnetic for all values of
the dilution $p$. This is the conventional paramagnetic phase that can be found for $h_x > h_\infty(\alpha)$ or, correspondingly,
for $\alpha < \alpha_\infty (h_x)$. Here, $h_{\infty}(\alpha)$ is the transverse field at which the undiluted bulk system undergoes the transition at fixed
$\alpha$ while $\alpha_{\infty}(h_x)$ is its critical dissipation strength at fixed $h_x$.

The behavior for $h_x < h_\infty(\alpha)$ [or $\alpha > \alpha_\infty (h_x)$] depends on the dilution $p$.
It is clear that magnetic long-range order is impossible for $p>p_c$, because the lattice consists of finite-size
clusters that are completely decoupled from each other. Each of these clusters acts as an independent magnetic moment.
For $h_x < h_\infty(\alpha)$ and $p>p_c$, the system is thus in a cluster super-paramagnetic phase.

Let us consider a single cluster of $s$ sites in more detail. For small transverse fields, its low-energy physics is equivalent
to that of a sub-Ohmic spin-boson model, i.e., a single effective Ising spin (whose moment is proportional to $s$)
in an effective transverse-field $h_x(s) \sim h_x e^{-Bs}$ with $B \sim \ln(J/h_x)$ and coupled  to a sub-Ohmic bath
with an effective dissipation strength $\alpha_s = s\alpha$.\cite{SenthilSachdev96,HoyosVojta06}
With increasing dissipation strength and/or decreasing transverse field, this sub-Ohmic spin-boson model
undergoes a quantum phase transition from a fluctuating to a localized (frozen) ground state.\cite{BullaTongVojta03}
This implies that sufficiently large percolation clusters are in the localized phase, i.e., they behave as classical
moments. The cluster super-paramagnetic phase thus consists of two regimes. If the transverse field is not too small,
$h_1(\alpha) < h_x < h_\infty(\alpha)$ [or if the dissipation is not too strong, $\alpha_1(h_x) > \alpha > \alpha_\infty(h_x)$],
static and dynamic clusters coexist.
Here, $h_1(\alpha)$ is the critical field of a \emph{single} spin in a bath of dissipation
strength $\alpha$ while $\alpha_1(h_x)$ is its critical dissipation strength in a given field $h_x$.
In contrast, for $h_x < h_1(\alpha)$ [or $\alpha > \alpha_1(h_x)$], all clusters are frozen,
and the system behaves purely classically.

Finally, for dilutions $p<p_c$, there is an infinite-spanning percolation cluster that can support
magnetic long-range order. Naively, one might expect
that the critical transverse-field (at fixed dissipation strength $\alpha$) decreases with dilution $p$ because the spins are missing neighbors.
However, in our case of sub-Ohmic dissipation, rare vacancy-free spatial regions can undergo the quantum phase transition
independently from the bulk system. As a consequence, the field-driven transition [transition (i) in Fig.\ \ref{fig:Ising_pd}]
is smeared,\cite{Vojta03a,HoyosVojta08}  and the ordered phase extends all the way to the clean critical field $h_\infty(\alpha)$
for all $p<p_c$. Analogous arguments apply to the critical dissipation strength at fixed transverse field $h_x$.

The infinite percolation cluster coexists with a spectrum of isolated finite-size clusters
whose behavior depends on the transverse field and dissipation
strength. Analogous to the super-paramagnetic phase discussed above, the ordered phase thus consists of two regimes.
For $h_1(\alpha) < h_x < h_\infty(\alpha)$ [or $\alpha_1(h_x) > \alpha > \alpha_\infty(h_x)$],
static (frozen) and dynamic clusters coexist with the long-range-ordered infinite cluster.
For $h_x < h_1(\alpha)$ [or $\alpha > \alpha_1(h_x)$], all clusters are frozen, and the system behaves
classically.

The phase diagram of the diluted quantum rotor model with sub-Ohmic dissipation (\ref{eq:rotoraction}) can be discussed along the
same lines. After fixing the bath
parameters $\zeta$ and $\omega_c$ and measuring all energies in terms of the exchange interaction $J$,
we are left with two parameters, the dilution $p$ and the dissipation strength $\alpha$. The zero-temperature behavior
of a single quantum rotor coupled to a sub-Ohmic bath is analogous to that of the corresponding quantum Ising spin.
With increasing dissipation strength, the rotor undergoes a quantum phase transition from a fluctuating to a localized
ground state. This follows, for instance, from mapping\cite{Sachdev_book99} the sub-Ohmic quantum rotor model onto a
one-dimensional classical Heisenberg chain with an interaction that falls off more slowly than $1/r^2$. This model
is known to have an ordered phase for sufficiently strong interactions.\cite{FILS78} As a result, all the arguments
used above to discuss the phase diagram of the diluted sub-Ohmic transverse-field Ising model carry over to the
rotor model Eq.~(\ref{eq:rotoraction}). The $\alpha$--$p$ phase diagram of the rotor model thus agrees with the
phase diagram shown in Fig.~\hyperref[fig:Ising_pd]{\ref{fig:Ising_pd}(a)}.

In the following sections, we investigate the percolation quantum phase transitions of the models Eqs.~(\ref{eq:Hamiltonian})
and (\ref{eq:rotoraction}), i.e., the transitions occurring when the dilution $p$ is tuned through the lattice percolation
threshold $p_c$. These transitions are marked in Fig.\ \ref{fig:Ising_pd} by (ii) and (iii).

\section{Diluted quantum rotor model in the large-$N$ limit}
\label{sec:large-N}

In this section, we focus on the diluted dissipative quantum rotor model in the
large-$N$ limit of an infinite number of order-parameter components. In this limit, the problem
turns into a self-consistent Gaussian model. Consequently, all calculations can be
performed explicitly.

\subsection{Single percolation cluster}
\label{subsec:single_cluster_observables}
We begin by considering a single percolation cluster of $s$ sites. For $\alpha>\alpha_\infty$, this cluster is locally in the ordered phase.
Following Refs.\ \onlinecite{VojtaSchmalian05b,AlAliVojta11}, it can therefore be described
 as a single large-$N$ rotor with moment $s$ coupled to a sub-Ohmic
dissipative bath of strength $\alpha_s=s\alpha$. Its effective action is given by
\begin{equation}
 \mathcal{A}_{{\rm eff}}=T\sum_{\omega_n}{\left[\frac{1}{2}\tilde{\psi}(\omega_n)\Gamma_n\tilde{\psi}(-\omega_n)-s\tilde{H}_z(\omega_n)\tilde{\psi}(-\omega_n)\right]}
\end{equation}
where $\Gamma_n=\epsilon+s \alpha \omega_c^{1-\zeta}|\omega_n|^{\zeta}$,
$\psi$ represents one rotor component and $H_z$ is an external field conjugate to the order parameter.

In the large-$N$ limit, the renormalized distance $\epsilon$ from criticality of the cluster is fixed by the large-$N$
 (spherical) constraint $\langle|\psi(\tau)|^2\rangle=1$. In terms of the Fourier transform, $\tilde\psi(\omega_n)$ defined by
\begin{equation}
 \psi(\tau)=T\sum_{\omega_n}{\tilde{\psi}(\omega_n)\exp{[-i\omega_n\tau]}},
\end{equation}
the large-$N$ constraint for a constant field $H_z$ becomes
\begin{equation}
T\sum_{\omega_n}{\frac{1}{\epsilon+s\alpha\omega_c^{1-\zeta}|\omega_n|^\zeta}}
+ \left (\frac{sH_z}{\epsilon} \right )^2 =1.
\label{Length-Const}
\end{equation}
Solving this equation gives the renormalized distance from criticality $\epsilon$ as a function of the cluster size $s$.

At zero temperature and field, the sum over the Matsubara frequencies turns into an integration, and the constraint equation reads
\begin{equation}
 \frac{1}{\pi}\int_0^{\omega_c}{{\rm d}\omega \frac{1}{\epsilon_0+s\alpha\omega_c^{1-\zeta}|\omega|^\zeta}=1}.
\label{Constraint-Eq}
\end{equation}
(We denote  the renormalized distance from criticality at zero temperature and field by $\epsilon_0$.)
The critical size $s_c$ above which the cluster freezes can be found by setting $\epsilon_0=0$
 and performing the integral (\ref{Constraint-Eq}). This gives
\begin{equation}
 s_c=1/\left[\pi\alpha(1-\zeta)\right].
\end{equation}

As we are interested in the critical behavior of the clusters, we now solve the constraint
 equation for cluster sizes close to the critical one, $s_c-s\ll s_c$.
This can be accomplished by subtracting the constraints at $s$ and $s_c$
 from each other.
We need to distinguish two cases: $1/2<\zeta<1$ and $\zeta < 1/2$.
 In the first case, the resulting integral can be easily evaluated after moving the cut-off $\omega_c$
to infinity. This gives
\begin{equation}
 \epsilon_0= \alpha s_c [ -\zeta\sin(\pi/\zeta) \alpha(s_c-s)]^{{\zeta}/{(1-\zeta)}}\omega_c~~(\mbox{for }\zeta>1/2).
\label{epsilon-zetagt}
\end{equation}
In the second case, $\zeta<1/2$, we can evaluate Eq.~(\ref{Constraint-Eq})
via a straight Taylor expansion in ($s_c-s$). This results in
\begin{equation}
 \epsilon_0 = \alpha^2 s_c \pi (1-2\zeta) (s_c-s)\omega_c~~~~(\mbox{for }\zeta<1/2).
\label{epsilon-zetalt}
\end{equation}
It will be useful to rewrite Eqs.~(\ref{epsilon-zetagt}) and (\ref{epsilon-zetalt}) in a more compact manner:
\begin{equation}
\epsilon_0 (s) =  [A_\zeta (1-s/s_c)]^{{x}/{(1-x)}}\omega_c,
\label{epsilon-zeta}
\end{equation}
where $A_\zeta=-(\alpha s_c)^{{1}/{\zeta}}\zeta \sin(\pi/\zeta) $ for $\zeta>1/2$, and $A_\zeta=(\alpha s_c)^{2}\pi (1-2\zeta)$ for $\zeta<1/2$,
and $x=\max\{1/2,\zeta \}$.

In order to compute thermodynamic quantities, we will also need the value of $\epsilon(s)$ at non zero temperature.
The constraint equation for small but nonzero temperature can be obtained by keeping the $\omega_n=0$ term
 in the frequency sum of Eq.~(\ref{Length-Const}) discrete, while representing all other modes in terms of an $\omega$-integral. This gives
\begin{equation}
 \frac{T}{\epsilon}+\frac{1}{\pi}\int_0^{\omega_c}{{\rm d}\omega \frac{1}{\epsilon+s\alpha \omega_c^{1-\zeta}|\omega|^\zeta}}=1.
\end{equation}
Solving this equation for asymptotically low temperatures results in the following behaviors.
For clusters larger than the critical size, $s>s_c$, $\epsilon$ vanishes linearly with $T$ via $\epsilon=Ts/(s-s_c)$. Clusters of exactly
the critical size have $\epsilon=A^x_\zeta \omega_c^{1-x}T^x$. For smaller clusters $(s<s_c)$, low temperatures
only lead to a small correction of the zero-temperature behavior $\epsilon_0$. Writing $\epsilon(T)=\epsilon_0+\delta T$, we obtain $\delta = [{s}/{(s_c-s)}][{x}/{(1-x)}]$. Clusters with sizes close to the critical one show a crossover from the off-critical
to the critical regime with increasing $T$.
For $s\lesssim s_c$, this means
\begin{equation}
 \epsilon(T)\approx
\begin{cases}
\epsilon_0(1+\delta T/\epsilon_0)  & (\mbox{for }\epsilon_0 \gg \epsilon_T), \\
\epsilon_T  & \text{(otherwise)},
\end{cases}
\label{eps_T}
\end{equation}
with $\epsilon_T = A_\zeta^x\omega_c^{1-x}T^x$.

The constraint equation at zero temperature but in a nonzero ordering field $H_z$ can be solved analogously. \cite{AlAliVojta11}
For asymptotically small fields, we find $\epsilon(H_z)=sH_z[s/(s-s_c)]^{1/2}$ in the case of clusters of size $s>s_c$.
At the critical size, $\epsilon(H_z)=[A_\zeta^x \omega_c^{1-x} (s_cH_z)^{2x}]^{1/(1+x)}$, and for $s<s_c$ we obtain
 $\epsilon(H_z)=\epsilon_0+\delta(sH_z)^2/\epsilon_0$.
Larger fields lead to a crossover from the off-critical to the critical regime. For $s\lesssim s_c$, it reads
\begin{equation}
 \epsilon(H_z)\approx
\begin{cases}
\epsilon_0[1+\delta(sH_z/\epsilon_0)^2]  & (\mbox{for }\epsilon_0 \gg \epsilon_{H_z}), \\
\epsilon_{H_z}  & \text{(otherwise)},
\end{cases}
\label{eps_Hz}
\end{equation}
with $\epsilon_{H_z} =[A_\zeta^x\omega_c^{1-x}(sH_z)^{2x}]^{1/(1+x)}$.

Observables of a single cluster can now be determined by taking the appropriate derivatives of the free energy $F_{cl}=-T\ln(Z)$ with
\begin{equation}
 Z=\prod_n {Z_n}
\end{equation}
 where
\begin{equation}
 Z_n=\frac{T}{\epsilon+s\alpha\omega_c^{1-\zeta}|\omega_n|^\zeta}\exp{\left(\frac{T}{2}
\frac{s\tilde{H}_z(\omega_n)s\tilde{H}_z(-\omega_n)}{\epsilon+s\alpha\omega_c^{1-\zeta}|\omega_n|^\zeta}\right)}.
\end{equation}
The dynamical (Matsubara) susceptibility and magnetization are then given by
\begin{equation}
 \chi_{cl}(i\omega_n)=\frac{s^2}{\epsilon+s\alpha \omega_c^{1-\zeta}|\omega_n|^\zeta}~,
\label{Susc-Matsub}
\end{equation}
and
\begin{equation}
 m_{cl}(\omega_n)=T\frac{s^2 \tilde H_z(\omega_n)}{\epsilon+s \alpha \omega_c^{1-\zeta}|\omega_n|^\zeta}~,
\end{equation}
respectively, where $\epsilon$ is given by the solution of constraint equation discussed above.
(Note that the contribution of a cluster of size $s$ to the uniform susceptibility is proportional to $s^2$).
Therefore, in the above two limiting cases, we can write the uniform and static susceptibility
 of a cluster of size $s<s_c$ as a function of  temperature as follows
\begin{equation}
\chi_{cl} (T) \approx
{s^2}/{\epsilon(T)}.
\label{chicluster}
 \end{equation}
Large clusters $(s>s_c)$ behave classically, $\chi_{cl}\approx s(s-s_c)/T$, at low-temperatures.
Finally, for the critical ones $\chi_{cl}\approx s^2/\epsilon_T$.

In order to calculate the retarded susceptibility $\chi_{cl}(\omega)$, we need to analytically continue the Matsubara susceptibility
  by performing a Wick rotation to real frequency, $i\omega_n\to \omega+i 0$.
The resulting dynamical susceptibility reads
\begin{equation}
 \chi_{cl}(\omega)=\frac{s^2}
{\epsilon+\alpha\omega_c^{1-\zeta}|\omega|^\zeta \left[\cos(\pi\zeta/2)-i \sin(\pi\zeta/2)\rm{sgn}(\omega)\right]}.
\label{chi:dynsusc}
\end{equation}

Using Eq.~(\ref{eps_Hz}), the single cluster magnetization in a small ordering constant field $H_z$ is given by
\begin{equation}
  m_{cl}=\chi_{cl}H_z\approx
\begin{cases}
{H_zs^2}/{\epsilon_0} & (\mbox{for }\epsilon_0\gg \epsilon_{H_z}), \\
 H_z s^2/\epsilon_{H_z} & \text{(otherwise)}.
\end{cases}
\label{sing-clus-mag}
\end{equation}

Thermal properties (at zero field) can be computed by using the  ``remarkable formulas'' derived by Ford {\it et al.},
\cite{FordLewisOConnell88}
 which express
 the free energy (the internal energy) of a quantum oscillator in a heat bath in terms of its
 susceptibility and the free energy  (internal energy) of the free oscillator.
For our model, they read, respectively
\begin{equation}
 F_{cl}=-\mu+\frac{1}{\pi}\int_0^\infty{{\rm d}\omega F_f(\omega,T) {\rm Im}{\left[\frac{d}{d\omega}\ln{\chi_{cl}(\omega)}\right]}},
\label{Eq:F_s}
\end{equation}
and
\begin{equation}
 U_{cl}=-\mu+\frac{1}{\pi}\int_0^\infty{{\rm d}\omega U_f(\omega,T) {\rm Im}{\left[\frac{d}{d\omega}\ln{\chi_{cl}(\omega)}\right]}}.
\label{Eq:U_s}
\end{equation}
Here, $F_f(\omega,T)=T\ln[2\sinh(\omega/(2T))]$ and $U_f(\omega,T)=(\omega/2)\coth(\omega/(2T))$.
 The extra $\mu$ terms stem from the Lagrange multiplier enforcing the large-$N$ constraint.\cite{AlAliVojta11}

The  entropy  $S_{cl}=(U_{cl}-F_{cl})/T$ can be calculated simply by inserting Eq.~(\ref{chi:dynsusc}) into
Eqs.~(\ref{Eq:F_s}) and (\ref{Eq:U_s}) and computing the resulting integral.
 For the dynamical clusters ($s<s_c$), the low-temperature entropy behaves as
\begin{equation}
 S_{cl}=B_\zeta \alpha s \omega_c^{1-\zeta}\frac{T^\zeta}{\epsilon_0},
\end{equation}
where $B_\zeta$ is a $\zeta$-dependent constant. At higher temperatures
(greater than $T^* \sim \epsilon_0^{1/\zeta}\omega_c^{1-1/\zeta}$),
the entropy becomes weakly dependent on $T$. \footnote{For $\zeta<1/2$ it has a logarithmic $T$-dependence, while for $\zeta>1/2$ its dependence on $T$ is even weaker.\cite{AlAliVojta11}}

In the low-$T$ limit, the specific heat $C_{cl}=T(\partial S_{cl}/\partial T)$ thus  behaves as
\begin{equation}
 C_{cl}=B_\zeta \zeta\alpha s \omega_c^{1-\zeta}\frac{T^\zeta}{\epsilon_0}.
\label{Ccl}
\end{equation}

\subsection{Complete system}
\label{subsec:complete_system}
After discussing the behavior of a single percolation-cluster, we now turn to the full diluted lattice model.
The low-energy density of states of  the dynamic clusters
$\rho_{dy}(\epsilon)=\sum_{s<s_c} {n_s \delta{(\epsilon - \epsilon_0(s))}}$  is obtained combining the single-cluster
result Eq.~(\ref{epsilon-zeta}) with the cluster-size distribution
Eq.~(\ref{eq:percscaling}), yielding
\begin{equation}
 \rho_{dy}(\epsilon) = A^{-1}_\zeta \left (x^{-1}-1\right ) \frac { n_{s(\epsilon)}s_c }{\omega_c }
\left(\frac{\epsilon}{\omega_c} \right)^{(1-2x)/{x}},
\label{DOS}
\end{equation}
where $s(\epsilon)$ is the size of a cluster with renormalized distance $\epsilon$ from criticality
 [which can be obtained inverting Eq.~(\ref{epsilon-zeta})]. Notice that $\rho_{dy}$ shows no dependence on $\epsilon$ in the case
$\zeta<1/2$. In particular, it does not diverge with $\epsilon\to 0$, in contrast to the case $\zeta>1/2$.

We now discuss the physics at the percolation transition, starting with the total magnetization $m$.
 We have to distinguish the contributions $m_{dy}$ from dynamical
 clusters, $m_{st}$ from  frozen finite-size clusters, and $m_{\infty}$ from the infinite percolation cluster, if any.
For zero ordering field $H_z$, $m_{dy}$ vanishes, because the dynamic clusters fluctuate between up and down.
 The frozen finite-size clusters individually have a non-zero magnetization,
but it sums up to zero ($m_{st}=0$), because they do not align coherently for $H_z=0$. Hence, the only coherent contribution
 to the total magnetization is $m_\infty$. Since the infinite cluster is long-range ordered for small transverse field $h_x<h_\infty(\alpha)$,
 its magnetization is proportional to the number $P_\infty$ of sites in the infinite cluster, giving
\begin{equation}
 m=m_\infty\sim P_\infty(p)\sim
\begin{cases}
|p-p_c|^{\beta_c}  & (\mbox{for }p<p_c), \\
0  & (\mbox{for }p>p_c).
\end{cases}
\label{Large_N_m}
\end{equation}
The magnetization critical exponent $\beta$ is therefore given by its classical lattice percolation value $\beta_c$.
In response to an infinitesimally small ordering field $H_z$, the frozen finite-size clusters align at zero temperature,
 leading to a jump in $m(H_z)$ at $H_z=0$. The magnitude of the jump is given by $m_{st}=\sum_{s>s_c}n_s$. At the percolation threshold,
 $m_{st}\approx (1-p_c)s_c^{2-\tau_c}$, and it vanishes exponentially for both $p\to 0$ and $p\to 1$. The total magnetization
 in an infinitesimal field (given by $m_\infty+m_{st}$) is analytic at $p=p_c$, and only clusters with sizes below $s_c$ are not polarized.

To estimate the contribution $m_{dy}$ of the dynamic clusters,
we integrate the magnetization of a single cluster Eq.~(\ref{sing-clus-mag}) over the DOS given in Eq.~(\ref{DOS}).
For $\zeta>1/2$, we find that
\begin{equation}
 m_{dy}=C_\zeta n_{s_c} s_c^2 \left (\frac{H_z s_c}{\omega_c}\right )^{3(1-\zeta)/(1+\zeta)},
\label{Dy_Mag}
\end{equation}
where $n_{s_c}$ is the density of critical clusters, and $C_\zeta = A_\zeta^{-3\zeta/(1+\zeta)}\zeta/(2\zeta-1)$.
For $\zeta<1/2$, the integration gives
\begin{equation}
 m_{dy}=\frac{n_{s_c}s_c^2}{A_\zeta} \left(\frac{s_c H_z}{\omega_c}\right)
\left[ 1+ \ln{\left(\frac{\theta_0}{(A_\zeta\omega_c s_c^2 H_z^2)^{1/3}}\right)}\right],
\end{equation}
where $\theta_0$ is a cut-off energy.

\begin{figure}
\includegraphics[height=5cm]{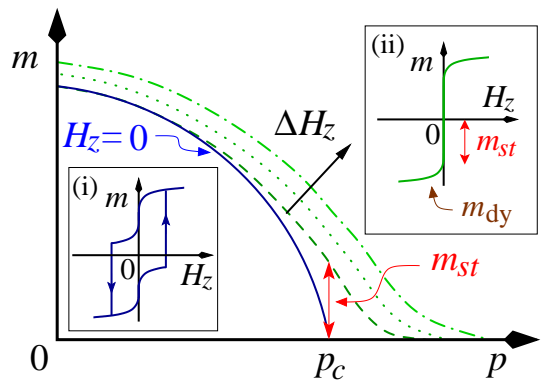}
\caption{(Color online) The magnetization as a function of dilution $p$ for different
 ordering field $H_z$ at absolute zero. The solid line is the magnetization at $H_z=0$
 (The contribution of the infinite cluster only). The dashed line is for an infinitesimal field and the remaining ones represents stronger fields.
Insets dysplay the histerisis curves in the (i) ordered and (ii) disordered phases.}
\label{mag}
\end{figure}

Because the three contributions to the magnetization
 have different field-dependence, the system shows unconventional hysteresis effects.
The infinite cluster has a regular hysteresis loop (for $p<p_c$), the finite-size frozen clusters do not show hysteresis, but they contribute
jumps in $m(H_z)$ at $H_z=0$, and the dynamic clusters contribute a continuous but singular term (see Fig. \ref{mag}).

The low-temperature susceptibility is dominated by the contribution $\chi_{st}$
of the static clusters, with each one adding a Curie term of the form $s(s-s_c)/T$. Summing over all static clusters,
close to the percolation threshold, we find that
\begin{equation}
 \chi_{st}\sim \sum_{s>s_c}{n_s\frac{s(s-s_c)}{T}}\sim\frac{1}{T}|p-p_c|^{-\gamma_c}.
\end{equation}
 For $p\to0$ and $p\to1$, the prefactor of the Curie term vanishes exponentially.
The infinite cluster contribution $\chi_\infty$ remains finite (per site) for $T\to0$, because the infinite cluster is in the ordered phase.

To determine the contribution $\chi_{dy}$ of the dynamical clusters, we integrate the single-cluster susceptibility
 Eq.~(\ref{chicluster}) over the low-energy DOS in Eq.~(\ref{DOS}). For $\zeta>1/2$, this gives
\begin{equation}
\chi_{dy}=C_\zeta^{\prime}\frac{ n_{s_c} s_c^3}{\omega_c} \left( \frac {T}{\omega_c} \right )^{1-2\zeta},
\label{Dy_Chi}
\end{equation}
with $C_\zeta^{\prime}=A^{-2\zeta}_\zeta [\zeta/(2\zeta-1)]$. For $\zeta<1/2$, we find
\begin{equation}
 \chi_{dy}=A_\zeta^{-1}\frac{n_{s_c}s_c^3}{\omega_c}
\left[ 1 + \ln \left(\frac{\theta_0}{(A_\zeta \omega_c T)^{1/2}} \right) \right].
\end{equation}

The retarded susceptibility of the fluctuating clusters can be obtained by integrating the single-cluster
 susceptibility Eq.~(\ref{chi:dynsusc}) over the distribution Eq.~(\ref{DOS}), this leads to
\begin{equation}
\textrm{Im}\, \chi_{dy}(\omega) =   D_\zeta
 \frac{ n_{s_c} s_c^3}{\omega_c} \left | \frac{\omega}{\omega_c}\right |^{1-2x} \textrm{sgn}(\omega),
\label{Imchi_dyzeta}
\end{equation}
with $D_\zeta=A^{-1}_\zeta(\frac{1}{x}\!-\!1)\pi \sin (\theta (\frac{1}{x}\!-\!2))/[\sin(\frac{\pi}{x})(\pi (1\!-\!\zeta))^{\frac{1}{x}-2}]$.
We notice that $\textrm{Im}\, \chi_{dy}$ has no $\omega$-dependence for $\zeta<1/2$.

Finally, we consider the heat capacity. The dynamical cluster contribution can be obtained by summing the single-cluster
heat capacity Eq.~(\ref{Ccl}) over $\rho_{dy}(\epsilon)$, yielding $C_{dy} \sim n_{s_c} s_c \left ({T}/{\omega_c} \right)^{1-\zeta}$
for $\zeta>1/2$ and $C_{dy} \sim n_{s_c} s_c \left ({T}/{\omega_c} \right)^{\zeta}$ for $\zeta<1/2$.

\section{Beyond the large-$N$ limit: scaling approach}
\label{sec:general}

In the last subsection, we have studied the percolation quantum phase transition
of the diluted sub-Ohmic rotor model Eq.~(\ref{eq:rotoraction})
in the large-$N$ limit.
Let us now go beyond the large-$N$ limit and consider the rotor model
with a finite number of components as well as the quantum
Ising model Eq.~(\ref{eq:Hamiltonian}).

We begin by analyzing a single percolation cluster of $s$ sites.
For strong dissipation $\alpha>\alpha_\infty$ (or weak fluctuations $h_x<h_\infty$), this cluster can be treated as
a compact object that fluctuates in (imaginary) time only.
As pointed out in Sec.~\ref{subsec:PD}, in the presence of sub-Ohmic dissipation,
such a cluster undergoes a continuous quantum phase transition from a fluctuating to
a localized phase as a function of increasing dissipation strength or, equivalently,
cluster size $s$.

Even though the critical behavior of this quantum phase transition is not exactly
solvable, we can still write down a scaling description of the cluster
free energy
\begin{equation}
F_{cl}(r,H_z,T) = b^{-1} F_{cl}(rb^{1/(\nu_s z_s)}, H_zb^{y_s}, Tb)
\label{eq:Fs}
\end{equation}
where $r=\alpha_s-\alpha_c =(s-s_c)\alpha$ is the distance from criticality, $b$ is
an arbitrary scale factor, and $\nu_s z_s$ and $y_s$ are the critical exponents of
the single-cluster quantum phase transition. (We use a subscript $s$ to distinguish
the single-cluster exponents from those associated with the percolation quantum
phase transition of the diluted lattice.)

Normally, one would expect the two exponents $\nu_s z_s$ and $y_s$ to be independent.
However, because the sub-Ohmic damping corresponds to a long-range interaction in
time, the exponent $\eta$ takes the mean-field value $2-\zeta$ for all
$\zeta$.\cite{FisherMaNickel72,Sak77,LuijtenBlote02} This also fixes the
exponent $y_s$ in Eq.~(\ref{eq:Fs}) to be $y_s=(1+\zeta)/2$.
Thus, there is only one independent
exponent in addition to $\zeta$; in the following we choose the susceptibility
exponent $\gamma_s$. This implies, via the usual scaling relations, that the correlation
time exponent is given by $\nu_s z_s = \gamma_s/\zeta$.

The values of the cluster exponents in the large-$N$ case of Sec.~\ref{sec:large-N}
are given by $\gamma_s=\zeta/(1-\zeta)$ and $\nu_s z_s = 1/(1-\zeta)$.
In the general case of finite-$N$ rotors and for the quantum Ising model,
they can be found numerically. Notice the scaling form of the free energy Eq.~(\ref{eq:Fs})
applies to bath exponents $\zeta>1/2$. For $\zeta<1/2$, the single-cluster
critical behavior is mean-field-like.

The behavior of single-cluster observables close to the (single-cluster) quantum
critical point can now be obtained by taking the appropriate derivatives of the
free energy Eq.~(\ref{eq:Fs}). For example, the static magnetic susceptibility
at $T=0$ and $H_z=0$ behaves as
\begin{equation}
\chi(r,\omega=0) \sim r^{-\gamma_s}~.
\end{equation}
Using this result, we can derive a
generalization of the probability distribution $\rho_{dy}(\epsilon)$ of the
inverse static susceptibilities $\epsilon =   \chi^{-1}$. We find
\begin{equation}
\rho_{dy}(\epsilon)= \int_1^{s_c} {\rm d}s\,n_s\, \delta\left[\epsilon - c (s_c-s)^{\gamma_s}\right] \sim n_{s_c}\,\epsilon^{(1-\gamma_s)/{\gamma_s}}
\label{eq:rho_general}
\end{equation}
right at the
percolation threshold. In the large-$N$ limit, $\gamma_s = \zeta/(1-\zeta)$ implying $\rho_{dy}(\epsilon) \sim \epsilon^{(1-2\zeta)/\zeta}$
in agreement with the explicit result in  Eq.~(\ref{DOS}).

Let us now discuss how the properties of the percolation quantum phase transition
in the general case differ from those obtained in the large-$N$ limit in Sec.\
\ref{subsec:complete_system}. We focus on the case $\zeta>1/2$. If the single-cluster critical behavior
is of mean-field type ($\zeta<1/2$), the functional forms of the results in Sec.\
\ref{subsec:complete_system} are not modified at all.
 The total magnetization is the sum of
the magnetization $m_\infty$ of the infinite percolation cluster, $m_{st}$ stemming from
the large ($s>s_c$) frozen percolation clusters, and $m_{dy}$ provided by the dynamic
clusters having $s<s_c$. Both $m_\infty$ and $m_{st}$ are completely independent
of the single-cluster critical behavior. The behavior of the spontaneous (zero-field)
magnetization across the percolation transition in the general case is thus identical
to that of the large-$N$ limit [see Eq.~(\ref{Large_N_m}) and Fig.~\ref{mag}]. In contrast, the
magnetization--magnetic field curve of the dynamic clusters does depend on the value of $\gamma_s$.
Integrating the single cluster-magnetization of all dynamic clusters [in analogy to Eq.~(\ref{sing-clus-mag})]
gives
\begin{equation}
m_{dy} \sim H_z^{[1-\zeta+2\zeta/\gamma_s]/(1+\zeta)}.
\label{eq:m_dyn_general}
\end{equation}
In the large-$N$ limit, this recovers the result Eq.~(\ref{Dy_Mag}), as expected.

The low-temperature susceptibility can be discussed along the same lines.
The contributions $\chi_\infty$ and $\chi_{st}$ do not depend on the single-cluster
critical behavior. Integrating the single-cluster susceptibility over all
dynamic clusters using (\ref{eq:rho_general}) yields (at $p=p_c$)
\begin{equation}
\chi_{dy} \sim T^{(1-\gamma_s)\zeta/\gamma_s}~.
\label{eq:chi_dyn_general}
\end{equation}
If we use the large-$N$ value of $\gamma_s$, we reproduce Eq.~(\ref{Dy_Chi}).

The scaling ansatz  Eq.~(\ref{eq:Fs}) for the single-cluster free energy thus allows us
to discuss the complete thermodynamics across the percolation quantum phase transition.
Dynamic quantities can be analyzed in the same manner. For example, the scaling form
of the single-cluster dynamic susceptibility reads
\begin{equation}
\chi_{cl}(r,H_z,T,\omega) = b^{2y_s-1} \chi_{cl}(rb^{1/(\nu_s z_s)}, H_zb^{y_s}, Tb, \omega b)
\label{eq:chis}
\end{equation}
The contribution of the fluctuating clusters to the low-temperature dynamic susceptibility
can be found by integrating the single-cluster contribution over the distribution
Eq.~(\ref{eq:rho_general}). This leads to
\begin{equation}
\textrm{Im}\, \chi_{dy}(\omega) \sim |\omega|^{(1-\gamma_s)\zeta/\gamma_s}\, \textrm{sgn}(\omega)~.
\label{eq:chi_omega_general}
\end{equation}
In the large-$N$ limit this corresponds to $\textrm{Im}\, \chi_{dy}(\omega) \sim |\omega|^{1-2\zeta}\, \textrm{sgn}(\omega)$
in agreement with Eq.~(\ref{Imchi_dyzeta}) for $\zeta>1/2$.

In summary, even though the critical behavior is not exactly solvable
for finite-$N$ rotors and quantum Ising models, we can express the properties of
the percolation quantum phase transition in terms of a single independent exponent
of the single-cluster problem (which can be found, e.g., numerically).

\section{Conclusions}
\label{sec:conclusions}

We have investigated the effects of local sub-Ohmic dissipation on the quantum phase transition
across the lattice percolation threshold of diluted quantum Ising and rotor models.
Experimentally, such local dissipation (with various
spectral densities) can be realized, e.g., in molecular magnets weakly coupled
to nuclear spins\cite{ProkofevStamp00,CWMBB00} or in magnetic nanoparticles in an insulating host.
\cite{Wernsdorfer01} Further potential applications include diluted two-level atoms in optical lattices coupled
to an electromagnetic field, random arrays of tunneling impurities in crystalline solids or, in the
future, large sets of coupled qubits in noisy environments.

As even a single spin or rotor undergoes a localization quantum phase
transition for sufficiently strong sub-Ohmic damping, the quantum dynamics of
large percolation clusters in the diluted lattice freezes completely.
The coexistence of these frozen clusters which effectively behave as classical
magnetic moments and smaller fluctuating clusters, if any, leads to unusual
properties of the percolation quantum phase transition. In this final section,
we put our results into broader perspective.

Let us compare the three different quantum phase transitions separating the paramagnetic
and ferromagnetic phases [transitions (i), (ii), and (iii) in Fig.~\ref{fig:Ising_pd}].
The generic transition (i) occurs as a function of transverse field or dissipation strength
for $p<p_c$. This transition is smeared by the mechanism of Ref.~\onlinecite{Vojta03a}
because rare vacancy-free spatial regions can undergo the quantum phase transition
independently from the bulk system. For $p<p_c$, these rare regions are weakly coupled
leading to magnetic long-range order instead of a quantum Griffiths phase.\cite{HoyosVojta08,HoyosVojta12}

In contrast, the percolation transitions (ii) and (iii) are not smeared but sharp.
The reason is that different percolation clusters are completely decoupled for $p>p_c$.
Thus, even if some of these clusters have undergone the (localization) quantum phase transition
and display local order, their local magnetizations do not align, leading to an incoherent
contribution to the global magnetization. Deviations from a pure percolation
scenario change this conclusion.  If the interaction has long-range tails (even very weak ones),
different frozen clusters will be coupled, and their magnetizations align coherently.
This leads to a smearing of the dilution-driven transition analogous to that of the transition (i).
However, if the long-range tail of the interaction is weak, the effects of the smearing
become important at the lowest energies only.
What is the difference between the percolation transitions (ii) and (iii) in Fig.\ \ref{fig:Ising_pd}?
If all percolation clusters are frozen [transitions (iii)]
low-temperature observables behave purely classically. If large frozen and smaller dynamic
clusters coexist [transitions (ii)] quantum fluctuations contribute to the
observables at the percolation transition.

We now compare the case of sub-Ohmic dissipation considered here to the cases of Ohmic and super-Ohmic
dissipation as well as the dissipationless case. To do so, we need to distinguish the quantum Ising model
and the rotor model.

The percolation transitions of the dissipationless and super-Ohmic rotor models
display conventional critical behavior, but with critical exponents that differ from the classical percolation
exponents.\cite{VojtaSchmalian05b} (This holds for the particle-hole symmetric case in which complex Berry phase
terms are absent from the action.\cite{FernandesSchmalian11}) In the Ohmic rotor model, the
percolation transition displays activated scaling as at infinite-randomness critical points.\cite{VojtaSchmalian05b}

For the diluted quantum Ising model, the percolation transition displays activated scaling already in the
dissipationless~\cite{SenthilSachdev96} and super-Ohmic cases.~\cite{HoyosVojta12} In the presence of Ohmic dissipation, sufficiently
large percolation clusters can undergo the localization transition independently from the bulk. The resulting percolation
transition\cite{HoyosVojta06} is similar to the one discussed in the present paper, it shows unusual properties due to an
interplay of frozen and dynamic percolation clusters.

All these results suggest that quantum phase transitions across the lattice percolation threshold can be classified
analogously to generic disordered phase transitions,\cite{VojtaSchmalian05,Vojta06} (provided the order parameter action does not contain complex terms).
If a single finite-size percolation cluster is below the lower critical dimension of the problem,
it can not undergo a phase transition independent of the bulk system. The resulting percolation transition
displays conventional critical behavior (this is the case for the dissipationless and super-Ohmic
rotor models). If a single finite-size cluster can undergo the transition by itself
(i.e., it is above the lower critical dimension of the problem), the resulting
percolation transition is unconventional with some observables behaving classically while others are
influenced by quantum fluctuations. This scenario applies to the sub-Ohmic models studied in this paper
as well as the Ohmic quantum Ising model. Finally, if a single percolation cluster is right at the
lower critical dimension (but does not undergo a phase transition), the percolation quantum phase transition
shows activated critical behavior. This scenario applies to the dissipationless quantum Ising model
as well as the Ohmic quantum rotor model.

\section*{Acknowledgements}

This work has been supported in part by the NSF under grant no. DMR-0906566, by FAPESP under Grant No. 2010/ 03749-4, and by CNPq
under grants No. 590093/2011-8 and No. 302301/2009-7.

\bibliographystyle{apsrev4-1}
\bibliography{rareregions}

\begin{thebibliography}{44}%
\makeatletter
\providecommand \@ifxundefined [1]{%
 \@ifx{#1\undefined}
}%
\providecommand \@ifnum [1]{%
 \ifnum #1\expandafter \@firstoftwo
 \else \expandafter \@secondoftwo
 \fi
}%
\providecommand \@ifx [1]{%
 \ifx #1\expandafter \@firstoftwo
 \else \expandafter \@secondoftwo
 \fi
}%
\providecommand \natexlab [1]{#1}%
\providecommand \enquote  [1]{``#1''}%
\providecommand \bibnamefont  [1]{#1}%
\providecommand \bibfnamefont [1]{#1}%
\providecommand \citenamefont [1]{#1}%
\providecommand \href@noop [0]{\@secondoftwo}%
\providecommand \href [0]{\begingroup \@sanitize@url \@href}%
\providecommand \@href[1]{\@@startlink{#1}\@@href}%
\providecommand \@@href[1]{\endgroup#1\@@endlink}%
\providecommand \@sanitize@url [0]{\catcode `\\12\catcode `\$12\catcode
  `\&12\catcode `\#12\catcode `\^12\catcode `\_12\catcode `\%12\relax}%
\providecommand \@@startlink[1]{}%
\providecommand \@@endlink[0]{}%
\providecommand \url  [0]{\begingroup\@sanitize@url \@url }%
\providecommand \@url [1]{\endgroup\@href {#1}{\urlprefix }}%
\providecommand \urlprefix  [0]{URL }%
\providecommand \Eprint [0]{\href }%
\providecommand \doibase [0]{http://dx.doi.org/}%
\providecommand \selectlanguage [0]{\@gobble}%
\providecommand \bibinfo  [0]{\@secondoftwo}%
\providecommand \bibfield  [0]{\@secondoftwo}%
\providecommand \translation [1]{[#1]}%
\providecommand \BibitemOpen [0]{}%
\providecommand \bibitemStop [0]{}%
\providecommand \bibitemNoStop [0]{.\EOS\space}%
\providecommand \EOS [0]{\spacefactor3000\relax}%
\providecommand \BibitemShut  [1]{\csname bibitem#1\endcsname}%
\let\auto@bib@innerbib\@empty
\bibitem [{\citenamefont {Thill}\ and\ \citenamefont
  {Huse}(1995)}]{ThillHuse95}%
  \BibitemOpen
  \bibfield  {author} {\bibinfo {author} {\bibfnamefont {M.}~\bibnamefont
  {Thill}}\ and\ \bibinfo {author} {\bibfnamefont {D.~A.}\ \bibnamefont
  {Huse}},\ }\href@noop {} {\bibfield  {journal} {\bibinfo  {journal} {Physica
  A}\ }\textbf {\bibinfo {volume} {214}},\ \bibinfo {pages} {321} (\bibinfo
  {year} {1995})}\BibitemShut {NoStop}%
\bibitem [{\citenamefont {Young}\ and\ \citenamefont
  {Rieger}(1996)}]{YoungRieger96}%
  \BibitemOpen
  \bibfield  {author} {\bibinfo {author} {\bibfnamefont {A.~P.}\ \bibnamefont
  {Young}}\ and\ \bibinfo {author} {\bibfnamefont {H.}~\bibnamefont {Rieger}},\
  }\href@noop {} {\bibfield  {journal} {\bibinfo  {journal} {Phys. Rev. B}\
  }\textbf {\bibinfo {volume} {53}},\ \bibinfo {pages} {8486} (\bibinfo {year}
  {1996})}\BibitemShut {NoStop}%
\bibitem [{\citenamefont {Fisher}(1992)}]{Fisher92}%
  \BibitemOpen
  \bibfield  {author} {\bibinfo {author} {\bibfnamefont {D.~S.}\ \bibnamefont
  {Fisher}},\ }\href@noop {} {\bibfield  {journal} {\bibinfo  {journal} {Phys.
  Rev. Lett.}\ }\textbf {\bibinfo {volume} {69}},\ \bibinfo {pages} {534}
  (\bibinfo {year} {1992})}\BibitemShut {NoStop}%
\bibitem [{\citenamefont {Fisher}(1995)}]{Fisher95}%
  \BibitemOpen
  \bibfield  {author} {\bibinfo {author} {\bibfnamefont {D.~S.}\ \bibnamefont
  {Fisher}},\ }\href@noop {} {\bibfield  {journal} {\bibinfo  {journal} {Phys.
  Rev. B}\ }\textbf {\bibinfo {volume} {51}},\ \bibinfo {pages} {6411}
  (\bibinfo {year} {1995})}\BibitemShut {NoStop}%
\bibitem [{\citenamefont {Vojta}(2006)}]{Vojta06}%
  \BibitemOpen
  \bibfield  {author} {\bibinfo {author} {\bibfnamefont {T.}~\bibnamefont
  {Vojta}},\ }\href@noop {} {\bibfield  {journal} {\bibinfo  {journal} {J.
  Phys. A}\ }\textbf {\bibinfo {volume} {39}},\ \bibinfo {pages} {R143}
  (\bibinfo {year} {2006})}\BibitemShut {NoStop}%
\bibitem [{\citenamefont {Vojta}(2010)}]{Vojta10}%
  \BibitemOpen
  \bibfield  {author} {\bibinfo {author} {\bibfnamefont {T.}~\bibnamefont
  {Vojta}},\ }\href@noop {} {\bibfield  {journal} {\bibinfo  {journal} {J. Low
  Temp. Phys.}\ }\textbf {\bibinfo {volume} {161}},\ \bibinfo {pages} {299}
  (\bibinfo {year} {2010})}\BibitemShut {NoStop}%
\bibitem [{\citenamefont {Vojta}\ and\ \citenamefont
  {Hoyos}(2008)}]{VojtaHoyos08b}%
  \BibitemOpen
  \bibfield  {author} {\bibinfo {author} {\bibfnamefont {T.}~\bibnamefont
  {Vojta}}\ and\ \bibinfo {author} {\bibfnamefont {J.~A.}\ \bibnamefont
  {Hoyos}},\ }in\ \href@noop {} {\emph {\bibinfo {booktitle} {Recent Progress
  in Many-Body Theories}}},\ \bibinfo {editor} {edited by\ \bibinfo {editor}
  {\bibfnamefont {J.}~\bibnamefont {Boronat}}, \bibinfo {editor} {\bibfnamefont
  {G.}~\bibnamefont {Astrakharchik}}, \ and\ \bibinfo {editor} {\bibfnamefont
  {F.}~\bibnamefont {Mazzanti}}}\ (\bibinfo  {publisher} {World Scientific},\
  \bibinfo {address} {Singapore},\ \bibinfo {year} {2008})\ p.\ \bibinfo
  {pages} {235}\BibitemShut {NoStop}%
\bibitem [{\citenamefont {Senthil}\ and\ \citenamefont
  {Sachdev}(1996)}]{SenthilSachdev96}%
  \BibitemOpen
  \bibfield  {author} {\bibinfo {author} {\bibfnamefont {T.}~\bibnamefont
  {Senthil}}\ and\ \bibinfo {author} {\bibfnamefont {S.}~\bibnamefont
  {Sachdev}},\ }\href@noop {} {\bibfield  {journal} {\bibinfo  {journal} {Phys.
  Rev. Lett.}\ }\textbf {\bibinfo {volume} {77}},\ \bibinfo {pages} {5292}
  (\bibinfo {year} {1996})}\BibitemShut {NoStop}%
\bibitem [{\citenamefont {Fernandes}\ and\ \citenamefont
  {Schmalian}(2011)}]{FernandesSchmalian11}%
  \BibitemOpen
  \bibfield  {author} {\bibinfo {author} {\bibfnamefont {R.~M.}\ \bibnamefont
  {Fernandes}}\ and\ \bibinfo {author} {\bibfnamefont {J.}~\bibnamefont
  {Schmalian}},\ }\href {\doibase 10.1103/PhysRevLett.106.067004} {\bibfield
  {journal} {\bibinfo  {journal} {Phys. Rev. Lett.}\ }\textbf {\bibinfo
  {volume} {106}},\ \bibinfo {pages} {067004} (\bibinfo {year}
  {2011})}\BibitemShut {NoStop}%
\bibitem [{\citenamefont {Vojta}\ and\ \citenamefont
  {Schmalian}(2005{\natexlab{a}})}]{VojtaSchmalian05}%
  \BibitemOpen
  \bibfield  {author} {\bibinfo {author} {\bibfnamefont {T.}~\bibnamefont
  {Vojta}}\ and\ \bibinfo {author} {\bibfnamefont {J.}~\bibnamefont
  {Schmalian}},\ }\href@noop {} {\bibfield  {journal} {\bibinfo  {journal}
  {Phys. Rev. B}\ }\textbf {\bibinfo {volume} {72}},\ \bibinfo {pages} {045438}
  (\bibinfo {year} {2005}{\natexlab{a}})}\BibitemShut {NoStop}%
\bibitem [{\citenamefont {Vojta}\ and\ \citenamefont
  {Sknepnek}(2006)}]{VojtaSknepnek06}%
  \BibitemOpen
  \bibfield  {author} {\bibinfo {author} {\bibfnamefont {T.}~\bibnamefont
  {Vojta}}\ and\ \bibinfo {author} {\bibfnamefont {R.}~\bibnamefont
  {Sknepnek}},\ }\href@noop {} {\bibfield  {journal} {\bibinfo  {journal}
  {Phys. Rev. B.}\ }\textbf {\bibinfo {volume} {74}},\ \bibinfo {pages}
  {094415} (\bibinfo {year} {2006})}\BibitemShut {NoStop}%
\bibitem [{\citenamefont {Wang}\ and\ \citenamefont
  {Sandvik}(2006)}]{WangSandvik06}%
  \BibitemOpen
  \bibfield  {author} {\bibinfo {author} {\bibfnamefont {L.}~\bibnamefont
  {Wang}}\ and\ \bibinfo {author} {\bibfnamefont {A.~W.}\ \bibnamefont
  {Sandvik}},\ }\href@noop {} {\bibfield  {journal} {\bibinfo  {journal} {Phys.
  Rev. Lett.}\ }\textbf {\bibinfo {volume} {97}},\ \bibinfo {pages} {117204}
  (\bibinfo {year} {2006})}\BibitemShut {NoStop}%
\bibitem [{\citenamefont {Wang}\ and\ \citenamefont
  {Sandvik}(2010)}]{WangSandvik10}%
  \BibitemOpen
  \bibfield  {author} {\bibinfo {author} {\bibfnamefont {L.}~\bibnamefont
  {Wang}}\ and\ \bibinfo {author} {\bibfnamefont {A.~W.}\ \bibnamefont
  {Sandvik}},\ }\href {\doibase 10.1103/PhysRevB.81.054417} {\bibfield
  {journal} {\bibinfo  {journal} {Phys. Rev. B}\ }\textbf {\bibinfo {volume}
  {81}},\ \bibinfo {pages} {054417} (\bibinfo {year} {2010})}\BibitemShut
  {NoStop}%
\bibitem [{\citenamefont {Millis}\ \emph {et~al.}(2001)\citenamefont {Millis},
  \citenamefont {Morr},\ and\ \citenamefont
  {Schmalian}}]{MillisMorrSchmalian01}%
  \BibitemOpen
  \bibfield  {author} {\bibinfo {author} {\bibfnamefont {A.~J.}\ \bibnamefont
  {Millis}}, \bibinfo {author} {\bibfnamefont {D.~K.}\ \bibnamefont {Morr}}, \
  and\ \bibinfo {author} {\bibfnamefont {J.}~\bibnamefont {Schmalian}},\
  }\href@noop {} {\bibfield  {journal} {\bibinfo  {journal} {Phys. Rev. Lett.}\
  }\textbf {\bibinfo {volume} {87}},\ \bibinfo {pages} {167202} (\bibinfo
  {year} {2001})}\BibitemShut {NoStop}%
\bibitem [{\citenamefont {Vojta}(2003)}]{Vojta03a}%
  \BibitemOpen
  \bibfield  {author} {\bibinfo {author} {\bibfnamefont {T.}~\bibnamefont
  {Vojta}},\ }\href@noop {} {\bibfield  {journal} {\bibinfo  {journal} {Phys.
  Rev. Lett.}\ }\textbf {\bibinfo {volume} {90}},\ \bibinfo {pages} {107202}
  (\bibinfo {year} {2003})}\BibitemShut {NoStop}%
\bibitem [{\citenamefont {Schehr}\ and\ \citenamefont
  {Rieger}(2006)}]{SchehrRieger06}%
  \BibitemOpen
  \bibfield  {author} {\bibinfo {author} {\bibfnamefont {G.}~\bibnamefont
  {Schehr}}\ and\ \bibinfo {author} {\bibfnamefont {H.}~\bibnamefont
  {Rieger}},\ }\href@noop {} {\bibfield  {journal} {\bibinfo  {journal} {Phys.
  Rev. Lett.}\ }\textbf {\bibinfo {volume} {96}},\ \bibinfo {pages} {227201}
  (\bibinfo {year} {2006})}\BibitemShut {NoStop}%
\bibitem [{\citenamefont {Schehr}\ and\ \citenamefont
  {Rieger}(2008)}]{SchehrRieger08}%
  \BibitemOpen
  \bibfield  {author} {\bibinfo {author} {\bibfnamefont {G.}~\bibnamefont
  {Schehr}}\ and\ \bibinfo {author} {\bibfnamefont {H.}~\bibnamefont
  {Rieger}},\ }\href@noop {} {\bibfield  {journal} {\bibinfo  {journal} {J.
  Stat. Mech.}\ ,\ \bibinfo {pages} {P04012}} (\bibinfo {year}
  {2008})}\BibitemShut {NoStop}%
\bibitem [{\citenamefont {Hoyos}\ and\ \citenamefont
  {Vojta}(2008)}]{HoyosVojta08}%
  \BibitemOpen
  \bibfield  {author} {\bibinfo {author} {\bibfnamefont {J.~A.}\ \bibnamefont
  {Hoyos}}\ and\ \bibinfo {author} {\bibfnamefont {T.}~\bibnamefont {Vojta}},\
  }\href@noop {} {\bibfield  {journal} {\bibinfo  {journal} {Phys. Rev. Lett.}\
  }\textbf {\bibinfo {volume} {100}},\ \bibinfo {pages} {240601} (\bibinfo
  {year} {2008})}\BibitemShut {NoStop}%
\bibitem [{\citenamefont {Hoyos}\ and\ \citenamefont
  {Vojta}(2012)}]{HoyosVojta12}%
  \BibitemOpen
  \bibfield  {author} {\bibinfo {author} {\bibfnamefont {J.~A.}\ \bibnamefont
  {Hoyos}}\ and\ \bibinfo {author} {\bibfnamefont {T.}~\bibnamefont {Vojta}},\
  }\href@noop {} {\bibfield  {journal} {\bibinfo  {journal} {Phys. Rev. B}\
  }\textbf {\bibinfo {volume} {85}},\ \bibinfo {pages} {174403} (\bibinfo
  {year} {2012})}\BibitemShut {NoStop}%
\bibitem [{\citenamefont {Hoyos}\ \emph {et~al.}(2007)\citenamefont {Hoyos},
  \citenamefont {Kotabage},\ and\ \citenamefont
  {Vojta}}]{HoyosKotabageVojta07}%
  \BibitemOpen
  \bibfield  {author} {\bibinfo {author} {\bibfnamefont {J.~A.}\ \bibnamefont
  {Hoyos}}, \bibinfo {author} {\bibfnamefont {C.}~\bibnamefont {Kotabage}}, \
  and\ \bibinfo {author} {\bibfnamefont {T.}~\bibnamefont {Vojta}},\
  }\href@noop {} {\bibfield  {journal} {\bibinfo  {journal} {Phys. Rev. Lett.}\
  }\textbf {\bibinfo {volume} {99}},\ \bibinfo {pages} {230601} (\bibinfo
  {year} {2007})}\BibitemShut {NoStop}%
\bibitem [{\citenamefont {Del~Maestro}\ \emph {et~al.}(2008)\citenamefont
  {Del~Maestro}, \citenamefont {Rosenow}, \citenamefont {M{\"u}ller},\ and\
  \citenamefont {Sachdev}}]{DRMS08}%
  \BibitemOpen
  \bibfield  {author} {\bibinfo {author} {\bibfnamefont {A.}~\bibnamefont
  {Del~Maestro}}, \bibinfo {author} {\bibfnamefont {B.}~\bibnamefont
  {Rosenow}}, \bibinfo {author} {\bibfnamefont {M.}~\bibnamefont {M{\"u}ller}},
  \ and\ \bibinfo {author} {\bibfnamefont {S.}~\bibnamefont {Sachdev}},\
  }\href@noop {} {\bibfield  {journal} {\bibinfo  {journal} {Phys. Rev. Lett.}\
  }\textbf {\bibinfo {volume} {101}},\ \bibinfo {pages} {035701} (\bibinfo
  {year} {2008})}\BibitemShut {NoStop}%
\bibitem [{\citenamefont {Vojta}\ \emph {et~al.}(2009)\citenamefont {Vojta},
  \citenamefont {Kotabage},\ and\ \citenamefont
  {Hoyos}}]{VojtaKotabageHoyos09}%
  \BibitemOpen
  \bibfield  {author} {\bibinfo {author} {\bibfnamefont {T.}~\bibnamefont
  {Vojta}}, \bibinfo {author} {\bibfnamefont {C.}~\bibnamefont {Kotabage}}, \
  and\ \bibinfo {author} {\bibfnamefont {J.~A.}\ \bibnamefont {Hoyos}},\
  }\href@noop {} {\bibfield  {journal} {\bibinfo  {journal} {Phys. Rev. B}\
  }\textbf {\bibinfo {volume} {79}},\ \bibinfo {pages} {024401} (\bibinfo
  {year} {2009})}\BibitemShut {NoStop}%
\bibitem [{\citenamefont {Vojta}\ \emph {et~al.}(2011)\citenamefont {Vojta},
  \citenamefont {Hoyos}, \citenamefont {Mohan},\ and\ \citenamefont
  {Narayanan}}]{VojtaHoyosMohanNarayanan11}%
  \BibitemOpen
  \bibfield  {author} {\bibinfo {author} {\bibfnamefont {T.}~\bibnamefont
  {Vojta}}, \bibinfo {author} {\bibfnamefont {J.~A.}\ \bibnamefont {Hoyos}},
  \bibinfo {author} {\bibfnamefont {P.}~\bibnamefont {Mohan}}, \ and\ \bibinfo
  {author} {\bibfnamefont {R.}~\bibnamefont {Narayanan}},\ }\href@noop {}
  {\bibfield  {journal} {\bibinfo  {journal} {Journal of Physics: Condensed
  Matter}\ }\textbf {\bibinfo {volume} {23}},\ \bibinfo {pages} {094206}
  (\bibinfo {year} {2011})}\BibitemShut {NoStop}%
\bibitem [{\citenamefont {Hoyos}\ and\ \citenamefont
  {Vojta}(2006)}]{HoyosVojta06}%
  \BibitemOpen
  \bibfield  {author} {\bibinfo {author} {\bibfnamefont {J.~A.}\ \bibnamefont
  {Hoyos}}\ and\ \bibinfo {author} {\bibfnamefont {T.}~\bibnamefont {Vojta}},\
  }\href@noop {} {\bibfield  {journal} {\bibinfo  {journal} {Phys. Rev. B}\
  }\textbf {\bibinfo {volume} {74}},\ \bibinfo {pages} {140401(R)} (\bibinfo
  {year} {2006})}\BibitemShut {NoStop}%
\bibitem [{\citenamefont {Bulla}\ \emph {et~al.}(2003)\citenamefont {Bulla},
  \citenamefont {Tong},\ and\ \citenamefont {Vojta}}]{BullaTongVojta03}%
  \BibitemOpen
  \bibfield  {author} {\bibinfo {author} {\bibfnamefont {R.}~\bibnamefont
  {Bulla}}, \bibinfo {author} {\bibfnamefont {N.-H.}\ \bibnamefont {Tong}}, \
  and\ \bibinfo {author} {\bibfnamefont {M.}~\bibnamefont {Vojta}},\ }\href
  {\doibase 10.1103/PhysRevLett.91.170601} {\bibfield  {journal} {\bibinfo
  {journal} {Phys. Rev. Lett.}\ }\textbf {\bibinfo {volume} {91}},\ \bibinfo
  {pages} {170601} (\bibinfo {year} {2003})}\BibitemShut {NoStop}%
\bibitem [{\citenamefont {Winter}\ \emph {et~al.}(2009)\citenamefont {Winter},
  \citenamefont {Rieger}, \citenamefont {Vojta},\ and\ \citenamefont
  {Bulla}}]{WRVB09}%
  \BibitemOpen
  \bibfield  {author} {\bibinfo {author} {\bibfnamefont {A.}~\bibnamefont
  {Winter}}, \bibinfo {author} {\bibfnamefont {H.}~\bibnamefont {Rieger}},
  \bibinfo {author} {\bibfnamefont {M.}~\bibnamefont {Vojta}}, \ and\ \bibinfo
  {author} {\bibfnamefont {R.}~\bibnamefont {Bulla}},\ }\href {\doibase
  10.1103/PhysRevLett.102.030601} {\bibfield  {journal} {\bibinfo  {journal}
  {Phys. Rev. Lett.}\ }\textbf {\bibinfo {volume} {102}},\ \bibinfo {pages}
  {030601} (\bibinfo {year} {2009})}\BibitemShut {NoStop}%
\bibitem [{\citenamefont {Harris}(1974)}]{Harris74b}%
  \BibitemOpen
  \bibfield  {author} {\bibinfo {author} {\bibfnamefont {A.~B.}\ \bibnamefont
  {Harris}},\ }\href@noop {} {\bibfield  {journal} {\bibinfo  {journal} {J.
  Phys. C}\ }\textbf {\bibinfo {volume} {7}},\ \bibinfo {pages} {3082}
  (\bibinfo {year} {1974})}\BibitemShut {NoStop}%
\bibitem [{\citenamefont {Stinchcombe}(1981)}]{Stinchcombe81}%
  \BibitemOpen
  \bibfield  {author} {\bibinfo {author} {\bibfnamefont {R.}~\bibnamefont
  {Stinchcombe}},\ }\href@noop {} {\bibfield  {journal} {\bibinfo  {journal}
  {J. Phys. C}\ }\textbf {\bibinfo {volume} {14}},\ \bibinfo {pages} {L263}
  (\bibinfo {year} {1981})}\BibitemShut {NoStop}%
\bibitem [{\citenamefont {dos Santos}(1982)}]{Santos82}%
  \BibitemOpen
  \bibfield  {author} {\bibinfo {author} {\bibfnamefont {R.~R.}\ \bibnamefont
  {dos Santos}},\ }\href@noop {} {\bibfield  {journal} {\bibinfo  {journal} {J.
  Phys. C}\ }\textbf {\bibinfo {volume} {15}},\ \bibinfo {pages} {3141}
  (\bibinfo {year} {1982})}\BibitemShut {NoStop}%
\bibitem [{\citenamefont {Cugliandolo}\ \emph {et~al.}(2005)\citenamefont
  {Cugliandolo}, \citenamefont {Lozano},\ and\ \citenamefont
  {Lozza}}]{CugliandoloLozanoLozza05}%
  \BibitemOpen
  \bibfield  {author} {\bibinfo {author} {\bibfnamefont {L.~F.}\ \bibnamefont
  {Cugliandolo}}, \bibinfo {author} {\bibfnamefont {G.~S.}\ \bibnamefont
  {Lozano}}, \ and\ \bibinfo {author} {\bibfnamefont {H.}~\bibnamefont
  {Lozza}},\ }\href@noop {} {\bibfield  {journal} {\bibinfo  {journal} {Phys.
  Rev. B}\ }\textbf {\bibinfo {volume} {71}},\ \bibinfo {pages} {224421}
  (\bibinfo {year} {2005})}\BibitemShut {NoStop}%
\bibitem [{\citenamefont {Prokofev}\ and\ \citenamefont
  {Stamp}(2000)}]{ProkofevStamp00}%
  \BibitemOpen
  \bibfield  {author} {\bibinfo {author} {\bibfnamefont {N.~V.}\ \bibnamefont
  {Prokofev}}\ and\ \bibinfo {author} {\bibfnamefont {P.~C.~E.}\ \bibnamefont
  {Stamp}},\ }\href@noop {} {\bibfield  {journal} {\bibinfo  {journal} {Rep.
  Progr. Phys.}\ }\textbf {\bibinfo {volume} {63}},\ \bibinfo {pages} {669}
  (\bibinfo {year} {2000})}\BibitemShut {NoStop}%
\bibitem [{\citenamefont {Chiorescu}\ \emph {et~al.}(2000)\citenamefont
  {Chiorescu}, \citenamefont {Wernsdorfer}, \citenamefont {M{\"u}ller},
  \citenamefont {B{\"o}gge},\ and\ \citenamefont {Barbara}}]{CWMBB00}%
  \BibitemOpen
  \bibfield  {author} {\bibinfo {author} {\bibfnamefont {I.}~\bibnamefont
  {Chiorescu}}, \bibinfo {author} {\bibfnamefont {W.}~\bibnamefont
  {Wernsdorfer}}, \bibinfo {author} {\bibfnamefont {A.}~\bibnamefont
  {M{\"u}ller}}, \bibinfo {author} {\bibfnamefont {H.}~\bibnamefont
  {B{\"o}gge}}, \ and\ \bibinfo {author} {\bibfnamefont {B.}~\bibnamefont
  {Barbara}},\ }\href@noop {} {\bibfield  {journal} {\bibinfo  {journal} {Phys.
  Rev. Lett.}\ }\textbf {\bibinfo {volume} {84}},\ \bibinfo {pages} {3454}
  (\bibinfo {year} {2000})}\BibitemShut {NoStop}%
\bibitem [{\citenamefont {Wernsdorfer}(2001)}]{Wernsdorfer01}%
  \BibitemOpen
  \bibfield  {author} {\bibinfo {author} {\bibfnamefont {W.}~\bibnamefont
  {Wernsdorfer}},\ }\href@noop {} {\bibfield  {journal} {\bibinfo  {journal}
  {Adv. Chem. Phys.}\ }\textbf {\bibinfo {volume} {118}},\ \bibinfo {pages}
  {99} (\bibinfo {year} {2001})}\BibitemShut {NoStop}%
\bibitem [{\citenamefont {Stauffer}\ and\ \citenamefont
  {Aharony}(1991)}]{StaufferAharony_book91}%
  \BibitemOpen
  \bibfield  {author} {\bibinfo {author} {\bibfnamefont {D.}~\bibnamefont
  {Stauffer}}\ and\ \bibinfo {author} {\bibfnamefont {A.}~\bibnamefont
  {Aharony}},\ }\href@noop {} {\emph {\bibinfo {title} {Introduction to
  Percolation Theory}}}\ (\bibinfo  {publisher} {CRC Press},\ \bibinfo
  {address} {Boca Raton},\ \bibinfo {year} {1991})\BibitemShut {NoStop}%
\bibitem [{Note1()}]{Note1}%
  \BibitemOpen
  \bibinfo {note} {In agreement with Subsec.\ \ref {subsec:models}, we define
  $p$ as the fraction of sites removed rather than the fraction of sites
  present.}\BibitemShut {Stop}%
\bibitem [{\citenamefont {Sachdev}(1999)}]{Sachdev_book99}%
  \BibitemOpen
  \bibfield  {author} {\bibinfo {author} {\bibfnamefont {S.}~\bibnamefont
  {Sachdev}},\ }\href@noop {} {\emph {\bibinfo {title} {Quantum phase
  transitions}}}\ (\bibinfo  {publisher} {Cambridge University Press},\
  \bibinfo {address} {Cambridge},\ \bibinfo {year} {1999})\BibitemShut
  {NoStop}%
\bibitem [{\citenamefont {Fr{\"o}hlich}\ \emph {et~al.}(1978)\citenamefont
  {Fr{\"o}hlich}, \citenamefont {Israel}, \citenamefont {Lieb},\ and\
  \citenamefont {Simon}}]{FILS78}%
  \BibitemOpen
  \bibfield  {author} {\bibinfo {author} {\bibfnamefont {J.}~\bibnamefont
  {Fr{\"o}hlich}}, \bibinfo {author} {\bibfnamefont {R.}~\bibnamefont
  {Israel}}, \bibinfo {author} {\bibfnamefont {E.~H.}\ \bibnamefont {Lieb}}, \
  and\ \bibinfo {author} {\bibfnamefont {B.}~\bibnamefont {Simon}},\ }\href
  {http://dx.doi.org/10.1007/BF01940327} {\bibfield  {journal} {\bibinfo
  {journal} {Commun. Math. Phys.}\ }\textbf {\bibinfo {volume} {62}},\ \bibinfo
  {pages} {1} (\bibinfo {year} {1978})}\BibitemShut {NoStop}%
\bibitem [{\citenamefont {Vojta}\ and\ \citenamefont
  {Schmalian}(2005{\natexlab{b}})}]{VojtaSchmalian05b}%
  \BibitemOpen
  \bibfield  {author} {\bibinfo {author} {\bibfnamefont {T.}~\bibnamefont
  {Vojta}}\ and\ \bibinfo {author} {\bibfnamefont {J.}~\bibnamefont
  {Schmalian}},\ }\href@noop {} {\bibfield  {journal} {\bibinfo  {journal}
  {Phys. Rev. Lett.}\ }\textbf {\bibinfo {volume} {95}},\ \bibinfo {pages}
  {237206} (\bibinfo {year} {2005}{\natexlab{b}})}\BibitemShut {NoStop}%
\bibitem [{\citenamefont {Al-Ali}\ and\ \citenamefont
  {Vojta}(2011)}]{AlAliVojta11}%
  \BibitemOpen
  \bibfield  {author} {\bibinfo {author} {\bibfnamefont {M.}~\bibnamefont
  {Al-Ali}}\ and\ \bibinfo {author} {\bibfnamefont {T.}~\bibnamefont {Vojta}},\
  }\href {\doibase 10.1103/PhysRevB.84.195136} {\bibfield  {journal} {\bibinfo
  {journal} {Phys. Rev. B}\ }\textbf {\bibinfo {volume} {84}},\ \bibinfo
  {pages} {195136} (\bibinfo {year} {2011})}\BibitemShut {NoStop}%
\bibitem [{\citenamefont {Ford}\ \emph {et~al.}(1988)\citenamefont {Ford},
  \citenamefont {Lewis},\ and\ \citenamefont
  {O'Connell}}]{FordLewisOConnell88}%
  \BibitemOpen
  \bibfield  {author} {\bibinfo {author} {\bibfnamefont {G.~W.}\ \bibnamefont
  {Ford}}, \bibinfo {author} {\bibfnamefont {J.~T.}\ \bibnamefont {Lewis}}, \
  and\ \bibinfo {author} {\bibfnamefont {R.~F.}\ \bibnamefont {O'Connell}},\
  }\href {\doibase 10.1103/PhysRevLett.55.2273} {\bibfield  {journal} {\bibinfo
   {journal} {Ann. Phys. (N.Y.)}\ }\textbf {\bibinfo {volume} {185}},\ \bibinfo
  {pages} {270} (\bibinfo {year} {1988})}\BibitemShut {NoStop}%
\bibitem [{Note2()}]{Note2}%
  \BibitemOpen
  \bibinfo {note} {For $\zeta <1/2$ it has a logarithmic $T$-dependence, while
  for $\zeta >1/2$ its dependence on $T$ is even weaker.\cite
  {AlAliVojta11}}\BibitemShut {NoStop}%
\bibitem [{\citenamefont {Fisher}\ \emph {et~al.}(1972)\citenamefont {Fisher},
  \citenamefont {Ma},\ and\ \citenamefont {Nickel}}]{FisherMaNickel72}%
  \BibitemOpen
  \bibfield  {author} {\bibinfo {author} {\bibfnamefont {M.~E.}\ \bibnamefont
  {Fisher}}, \bibinfo {author} {\bibfnamefont {S.-K.}\ \bibnamefont {Ma}}, \
  and\ \bibinfo {author} {\bibfnamefont {B.~G.}\ \bibnamefont {Nickel}},\
  }\href@noop {} {\bibfield  {journal} {\bibinfo  {journal} {Phys. Rev. Lett.}\
  }\textbf {\bibinfo {volume} {29}},\ \bibinfo {pages} {917} (\bibinfo {year}
  {1972})}\BibitemShut {NoStop}%
\bibitem [{\citenamefont {Sak}(1977)}]{Sak77}%
  \BibitemOpen
  \bibfield  {author} {\bibinfo {author} {\bibfnamefont {J.}~\bibnamefont
  {Sak}},\ }\href {\doibase 10.1103/PhysRevB.15.4344} {\bibfield  {journal}
  {\bibinfo  {journal} {Phys. Rev. B}\ }\textbf {\bibinfo {volume} {15}},\
  \bibinfo {pages} {4344} (\bibinfo {year} {1977})}\BibitemShut {NoStop}%
\bibitem [{\citenamefont {Luijten}\ and\ \citenamefont
  {Bl{\"o}te}(2002)}]{LuijtenBlote02}%
  \BibitemOpen
  \bibfield  {author} {\bibinfo {author} {\bibfnamefont {E.}~\bibnamefont
  {Luijten}}\ and\ \bibinfo {author} {\bibfnamefont {H.~W.~J.}\ \bibnamefont
  {Bl{\"o}te}},\ }\href@noop {} {\bibfield  {journal} {\bibinfo  {journal}
  {Phys. Rev. Lett}\ }\textbf {\bibinfo {volume} {89}},\ \bibinfo {pages}
  {025703} (\bibinfo {year} {2002})}\BibitemShut {NoStop}%
\end{thebibliography}%

\end{document}